\documentclass[epj,nopacs]{svjour}
\usepackage{graphicx,amsmath,amssymb,ifthen}

\newcommand{\Kkappa}[2]{\mathcal{L}_{#1}(\{#2\})}
\newcommand{\KK}[2]{\mathcal{K}_{#1}(\{#2\})}
\begin{document}
\title{Transition rates via Bethe ansatz for the spin-$\frac{1}{2}$ Heisenberg
  chain}
\authorrunning{Biegel, Karbach and Muller}
\author{Daniel Biegel\inst{1}
  \and Michael Karbach\inst{1} 
  \and Gerhard M{\"u}ller\inst{2}}
\institute{
  Bergische Universit{\"a}t Wuppertal, 
  Fachbereich Physik,
  D-42097 Wuppertal, Germany,
  \email{michael@karbach.org}\\  
  Department of Physics,
  University of Rhode Island, 
  Kingston RI 02881-0817, USA,
  \email{gmuller@uri.edu}}
\date{\today}
\abstract{We use the exact determinantal representation derived by Kitanine,
  Maillet, and Terras for matrix elements of local spin operators between Bethe
  wave functions of the one-dimensional $s=\frac{1}{2}$ Heisenberg model to
  calculate and numerically evaluate transition rates pertaining to dynamic spin
  structure factors. For real solutions $z_1,\ldots,z_r$ of the Bethe ansatz
  equations, the size of the determinants is of order $r\times r$.  We present
  applications to the zero-temperature spin fluctuations parallel and
  perpendicular to an external magnetic field.}
\maketitle
%
\section{Introduction}
\label{sec:Introduction}
%
The one-dimensional (1D) $s=\frac{1}{2}$ Heisenberg model was solved, in
principle, by Bethe some 70 years ago using an ad-hoc trial wave function, now
famously known as the Bethe ansatz \cite{Beth31}. Decades went by before the
power and scope of this method of exact analysis became widely known for
applications to spectrum and thermodynamics of a select class of completely
integrable model systems \cite{KBI93}.

Until recently, the most vexing exception to immense progress in the further
development of the Bethe ansatz has been the absence of a practical method to
use the exactly known and readily available Bethe wave functions for the
explicit calculation of transition matrix elements. The knowledge of such
transition rates is of paramount importance for an understanding of dynamic
correlation functions in relation to the underlying quasiparticles and for the
interpretation of experimental probes of quantum fluctuations in quasi-1D
magnetic compounds.

It was most remarkable, therefore, when Kitanine, Maillet, and Terras
\cite{KMT99} succeeded in reducing matrix elements between Bethe wave functions
for local spin operators to determinantal expressions. Here we use these
expressions and the norms previously determined by Korepin \cite{Kore86} to
calculate dynamic spin structure factors of the Heisenberg antiferromagnet with
periodic boundary conditions and a magnetic field:
\begin{equation}\label{eq:Hh}
  H = \sum_{n=1}^N \left[J{\bf S}_n \cdot {\bf S}_{n+1}  - hS_n^z\right].
\end{equation}

%
\section{Transition rates expressed as determinants}
\label{sec:Determinants}
%
Consider Bethe wave functions with $z$-component $S_T^z=N/2-r$ of the total
spin. They are specified by sets of rapidities $z_1,\ldots,z_r$, which are
solutions of the Bethe ansatz equations
\begin{equation}
\label{eq:bae}
N\phi(z_i) = 2\pi I_i + \sum_{j\neq i}\phi\bigl [(z_i-z_j)/2\bigr ],
\quad i=1,\ldots,r,
\end{equation}
where $\phi(z) \equiv 2\arctan z$. The Bethe quantum numbers $I_i$ provide a natural
classification of the spectrum. The challenge is to calculate transition rates
for physically motivated operators from the information encoded in the $z_i$.
Using the Bethe wave function directly is feasible but computationally
inefficient and therefore limited to relatively small systems $(r\lesssim12)$
\cite{KBM02}. The combined advances reported in Refs.~\cite{KMT99,Kore86} are
the basis of a much more powerful approach, which will be used in the following
to calculate transition rates
\begin{equation}
  \label{eq:trarat}
  M_{\lambda}^{\mu}(q) \equiv \frac{|\langle\psi_0|S^{\mu}_q|\psi_\lambda\rangle|^2}{\|\psi_0\|^{2}\|\psi_\lambda\|^{2}},\quad \mu=z,+,-
\end{equation} 
from the ground state of (\ref{eq:Hh}) for the operators
\begin{equation}\label{eq:spiflu}
S_q^\mu = \frac{1}{\sqrt{N}}\sum_n\,e^{iqn}S_n^\mu,\quad \mu=z,+,-.
\end{equation}
They probe the parallel $(\mu=z)$ and the perpendicular $(\mu=+,-)$ spin
fluctuations at zero temperature. 

The transition rates for the parallel spin fluctuations as inferred from
Eq.~(5.12) of Ref.~\cite{KMT99} for the matrix element
$\langle\psi_0|S_n^z|\psi_{\lambda}\rangle$ can be brought into the form
\begin{equation}
  \label{eq:3}
  M_{\lambda}^{z}(q) = \frac{N}{4} \frac{|\Omega^{z}|^2}{\|\psi_0\|^{2}\|\psi_\lambda\|^{2}},
\end{equation}
where
\begin{eqnarray}
  \label{eq:4}
  \Omega^{z}(\{z^{0}_j\}_{r},\{z_j\}_{r}) &=& 
  \prod_{j=1}^r \frac{z^{0}_j + i}{z_j + i} \nonumber \\
  &\times& \frac{2^r \det(\mathsf{\bar H}-2\mathsf{\bar P})}%
           {\prod\limits_{i<j}^{r} (z_i^{0} - z_j^{0}) \prod\limits_{i > j}^{r} (z_{i} - z_{j})},
\end{eqnarray}
\begin{eqnarray}\label{eq:5}
  \mathsf{\bar H}_{ab} &=& \frac{i}{z^{0}_a - z_b} \nonumber \\
   && \hspace*{-15mm} \times \left(
     \prod_{j \neq a}^{r} (z_j^0 - z_b + 2i) - d(z_b) \prod_{j \neq a}^{r} (z_j^0 - z_b - 2i)
   \right), \\
   \mathsf{\bar P}_{ab} &=& \frac{i}{{(z_a^0)}^2+1} \prod_{j=1}^r (z_j - z_b +2i), \label{eq:7}
 \end{eqnarray}
 \begin{equation}
  \label{eq:8}
  d(z_i) = \left( \frac{z_i-i}{z_i+i} \right)^N.
\end{equation}
The rapidities $\{z_{i}^{0}\}$ and $\{z_{i}\}$ belong to the ground state
$|\psi_0\rangle$ and to the excited state $|\psi_\lambda\rangle$, respectively.  The norm of
$|\psi_\lambda\rangle$ from Ref.~\cite{Kore86}, transcribed to our notation, reads:
\begin{equation}\label{eq:9}
  \|\psi_\lambda\|^{2} =  
  \left( \prod_{i \neq j}^{r} \frac{z_{i} - z_{j} - 2i}{z_{i} - z_{j}} \right) 
  \det\mathsf{N}(\{z_i\}),
\end{equation}
\begin{equation}\label{eq:10}
  \mathsf{N}_{ab} = 
  -2i \frac{\partial}{\partial z_b} \ln 
  \left\{ \left( \frac{z_a +i}{z_a -i} \right)^N 
    \prod_{k\neq a}^{r} \frac{z_a - z_k - 2i}{z_a - z_k +2i} \right\},
\end{equation}
and can be compactified into
\begin{equation}
  \label{eq:11}
  \|\psi_\lambda\|^{2} =  \frac{2^{r^{2}}\det \mathsf{K}(\{z_{i}\})}%
  {\prod\limits_{i < j}^{r} K(z_{i}-z_{j})\prod\limits_{i < j}^{r}(z_{i}-z_{j})^{2}},
\end{equation}
where
\begin{equation}
  \label{eq:12}
  \mathsf{K}_{ab}= 
  \begin{cases}
    K(z_{a}-z_{b})                                               &: a\neq b \\ 
    N\kappa(z_{a})-\sum_{j\neq a}^{r}K(z_{a}-z_{j}) &: a=b
  \end{cases}
\end{equation}
\begin{equation}\label{eq:34}
  \kappa(z) \doteq \frac{2}{1+z^{2}}, \quad
  K(z) \doteq \frac{4}{4+z^{2}}.
\end{equation}
For real $z_i$ the absolute square of (\ref{eq:4}) becomes:
\begin{equation}
  \label{eq:13}
  |\Omega^{z}|^{2} = 
  \frac{ 2^{2r}\prod\limits_{i=1}^{r}[\kappa(z_{i})/\kappa(z_{i}^{0})]}%
   {\prod\limits_{i<j}^{r}(z_{i}^{0}-z_{j}^{0})^{2}\prod\limits_{i<j}^{r}(z_{i}-z_{j})^{2}}
   |\det(\mathsf{\bar H}-2\mathsf{\bar P})|^{2}, 
\end{equation}
Rescaling $\mathsf{\bar H,\bar P}$ consolidates powers of
two. The calculation of the transition rate is thus reduced to the evaluation
of three $r\times r$ determinants:
\begin{eqnarray}
  \label{eq:21}
  M_{\lambda}^{z}(q) &=&
  \frac{N}{4}\frac{\Kkappa{r}{z_{i}}}{\Kkappa{r}{z_{i}^{0}}}
   \KK{r}{z_{i}^{0}} \KK{r}{z_{i}}
   \nonumber \\ && \times 
   \frac{|\det(\mathsf{H}-\mathsf{P})|^{2}}%
        {\det \mathsf{K}(\{z_{i}\})\det\mathsf{K}(\{z_{i}^{0}\})},
\end{eqnarray}
\begin{equation}
  \label{eq:19}
  \Kkappa{r}{z_{i}} \doteq \prod_{i=1}^{r}\kappa(z_{i}), \quad
  \KK{r}{z_{i}} \doteq \prod_{i<j}^{r}K(z_{i}-z_{j}),
\end{equation}
\begin{eqnarray}
  \label{eq:15}
  \mathsf{H}_{ab} &\doteq & \frac{i}{z^{0}_a - z_b} \nonumber \\
 && \hspace*{-0.9cm} \times \left(
     \prod_{j \neq a}^{r} G(z_j^0 - z_b) - d(z_b) \prod_{j \neq a}^{r} G^{*}(z_j^0 - z_b)
   \right), \\
   \mathsf{P}_{ab} &\doteq& i2\kappa(z_a^0) \prod_{j=1}^r G(z_j - z_b),
 \end{eqnarray}
\begin{equation}
  \label{eq:17}
  G(z) \doteq \frac{z}{2} + i.
\end{equation}
Along similar lines of manipulation, the transition rates 
\begin{equation}
  \label{eq:22}
   M_{\lambda}^{\pm}(q) = N \frac{|\Omega^\pm|^2}{\|\psi_0\|^{2} \|\psi_\lambda\|^{2}},
\end{equation}
for the perpendicular spin fluctuations as inferred from the Eqs. (5.3) and
(5.7) of Ref.~\cite{KMT99} for the matrix element $\langle\psi_0|S_n^+|\psi_{\lambda}\rangle$ are
brought into the form:
\begin{eqnarray}
  \label{eq:25}
     M_{\lambda}^{\pm}(q) &=&
  \left(\frac{\Kkappa{r}{z_{i}^{0}}}{\Kkappa{r\pm 1}{z_{i}}}\right)^{\pm1}
   \KK{r}{z_{i}^{0}}\KK{r\pm1}{z_{i}} 
   \nonumber \\ && \hspace*{10mm}\times 
   \frac{N |\det \mathsf{H^{\pm}}|^{2}}%
        {\det \mathsf{K}(\{z_{i}\})\det\mathsf{K}(\{z_{i}^{0}\})},
\end{eqnarray}
\begin{eqnarray} 
  \mathsf{H}^+_{ab} &=& \frac{i}{z_a - z^0_b} 
  \nonumber \\ &\times&
  \left(\prod_{j \neq a}^{r+1} G(z_j - z^0_b) - d(z^0_b) 
    \prod_{j \neq a}^{r+1} G^{*}(z_j - z^0_b) \right) 
  \nonumber \\
  &&\nonumber\\
  \mathsf{H}^+_{a,r+1} &=& i\kappa(z_{a}), 
  \quad a=1,\ldots r+1, \; b=1,\ldots,r,
\end{eqnarray}
\begin{eqnarray} 
  \mathsf{H}^-_{ab} &=& \frac{i}{z_a^{0} - z_b} 
  \nonumber \\ &\times&
  \left(\prod_{j \neq a}^{r} G(z_j^{0} - z_b) - d(z_b) 
    \prod_{j \neq a}^{r} G^{*}(z_j^{0} - z_b) \right) 
  \nonumber \\
  &&\nonumber\\
  \mathsf{H}^-_{ar} &=& i\kappa(z_a^0), 
  \quad a=1,\ldots r, \; b=1,\ldots,r.
\end{eqnarray}

%
\section{Applications}
\label{sec:Applications}
\begin{figure}[ht]
  \centering
  \includegraphics[width=60mm,angle=-90]{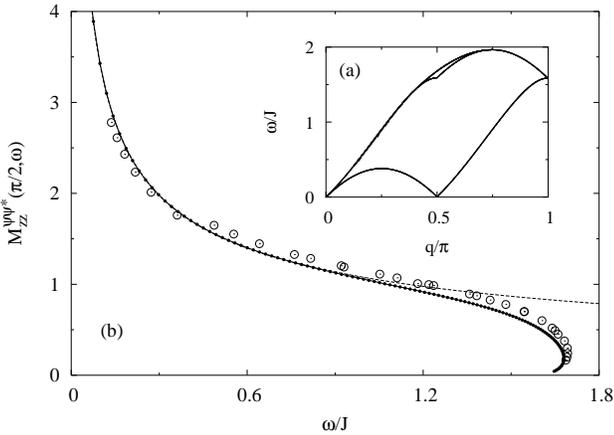}

\caption{(a) Energy versus wave number of the $\psi\psi^*$ excitations for $N\to\infty$. (b) Scaled
  transition rates between the ground state and the $\psi\psi^*$ states at $q=\pi/2$
  for $N=12,16,\ldots,32$ (open circles) and $N=512$ (closed circles). The dashed
  line is a fit $a+b\omega^{\eta-2}$ of the $N=512$ data at $\omega/J\leq 0.5$. All results are
  for $M_z=N/4$.}
\label{fig:fig1}
\end{figure}

%
To demonstrate the efficacy of expressions (\ref{eq:21}) and (\ref{eq:25}) for
the transition rates $M_\lambda^\mu(q),~\mu=z,+,-$, we present three applications to the
dynamic structure factors
\begin{equation}\label{eq:dssf}
S_{\mu\bar{\mu}}(q,\omega) = 2\pi\sum_\lambda M_\lambda^\mu(q)\delta\left(\omega-\omega_\lambda\right).
\end{equation}
We begin with the parallel spin fluctuations at magnetization
$M_z/N=\frac{1}{4}$ (half the saturation value). The spectral weight of
$S_{zz}(q,\omega)$ was shown to be dominated by a set of collective excitations
consisting of two unbound quasiparticles named $\psi$ and $\psi^*$ \cite{KM00}. The
spectral range of the $\psi\psi^*$ continuum, shown in Fig.~\ref{fig:fig1}(a), has soft modes at
$q=0, \pi/2$ and is partly folded back onto itself along a stretch of its upper
boundary.

In Fig.~\ref{fig:fig1}(b) we have plotted the scaled transition rates
$M_{zz}^{\psi\psi^*}(\pi/2,\omega)=NM_\lambda^z(\pi/2)$ versus $\omega$ pertaining to the $\psi\psi^*$
excitations. The open circles are results previously obtained for $N\leq32$ by
calculating matrix elements directly from Bethe wave functions \cite{KBM02}.
The full circles are data for $N=512$ as obtained from the determinantal
expression (\ref{eq:21}).

The function $M_{zz}^{\psi\psi^*}(\pi/2,\omega)$ varies smoothly across the entire $\psi\psi^*$
continuum including the fold. The observed infrared singularity,
$M_{zz}^{\psi\psi^*}(\pi/2,\omega) \sim \omega^{\eta-2}$, $\eta-2=-0.468\ldots$ confirms exact predictions
\cite{Hald80,FGM+96}. At the other end of the continuum, $M_{zz}^{\psi\psi^*}(\pi/2,\omega)$
tends to go to zero or a value close to zero.

\begin{figure}[t]
  \centering
  \includegraphics[width=60mm,angle=-90]{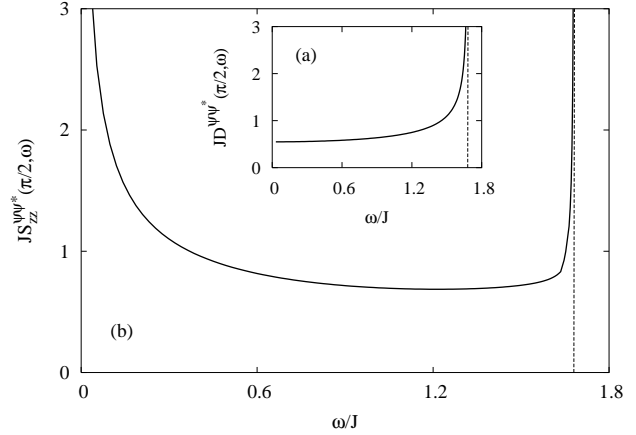}

\caption{(a) Density of $\psi\psi^*$ states and (b) spectral weight distribution of
  the $\psi\psi^*$ states in $S_{zz}(q,\omega)$ at $q=\pi/2$ and $M_z=N/4$ from data for $N=512$.}
\label{fig:fig2}
\end{figure}

The density $D^{\psi\psi^*}(\pi/2,\omega) =2\pi/[N(\omega_{i+1}-\omega_i)]$ of $\psi\psi^*$ states is
shown in Fig.~\ref{fig:fig2}(a). It is flat and featureless except near the
upper band edge, where the fold in the continuum produces a square-root
divergence \cite{KM00}. The spectral-weight distribution
$S_{zz}^{\psi\psi^*}(\pi/2,\omega)$ resulting from the product of $M_{zz}^{\psi\psi^*}$ and
$D^{\psi\psi^*}$ is then a double-peak structure as shown in Fig.~\ref{fig:fig2}(b)
with the divergences at the lower and upper band edges caused by the transition
rates and the density of states, respectively. At frequencies where the
continuum is folded back, the spectral weight of two lines of excitations
must be added up to produce the correct lineshape, causing a (barely visible)
singularity within the band.

\begin{figure}[ht]
  \centering
  \includegraphics[width=58mm,angle=-90]{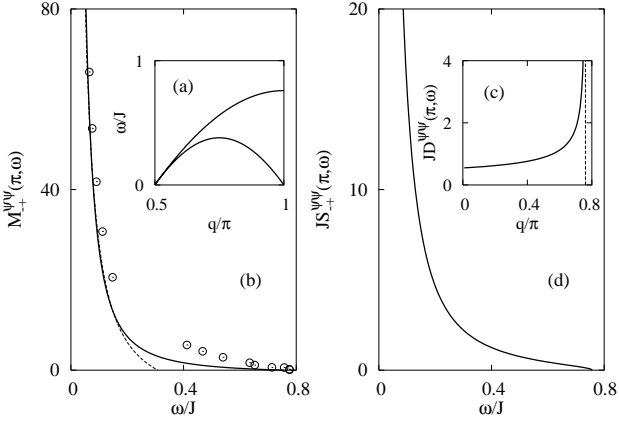}

\caption{(a) Continuum of $\psi\psi$ excitations. (b) Scaled transition
  rates between the ground state and the $\psi\psi$ states at $q=\pi$ for
  $N=12,16,\ldots,28$ (open circles) and $N=1536$ (solid line). The dashed line is a
  fit $a+b\omega^{1/ \eta-2}$ of the $N=1536$ data at $\omega\leq0.25$. (c) Density of $\psi\psi$
    states at $q=\pi$ for $N=1536$. (d) Lineshape at $q=\pi$ of the $\psi\psi$
    contribution to $S_{- +}(q,\omega)$. All results are for $M_z=N/4$.}
\label{fig:fig3}
\end{figure}

Now consider the perpendicular spin fluctuations as described by $S_{-
  +}(q,\omega)$, again at $M_z/N=\frac{1}{4}$. At the zone boundary $(q=\pi)$, some
98\% of the spectral weight is carried by a continuum of collective excitations
consisting of two $\psi$ quasiparticles \cite{KBM02}. The shape of the $\psi\psi$
continuum is shown in Fig.~\ref{fig:fig3}(a) The scaled transition rates $M_{-
  +}^{\psi\psi}(\pi,\omega)=NM_\lambda^-(\pi)$ pertaining to the $\psi\psi$ excitations are plotted versus
$\omega$ in Fig.~\ref{fig:fig3}(b) for $N=1536$ by numerical evaluation of the
determinantal expression (\ref{eq:25}). Also shown are data points for
$N=12,16,\ldots,28$ previously obtained by a different method \cite{KBM02}.

Again we find a smooth variation of the transition rates across the continuum
with distinct endpoint singularities. The infrared divergence is strong, $S_{-
  +}^{\psi\psi}(\pi,\omega) \sim\omega^{1/ \eta-2}, \quad 1/ \eta-2=-1.346\ldots$ as predicted
\cite{Hald80,FGM+96}. At the upper band edge, the transition rates approach zero
in what looks like a linear trend. The density $D^{\psi\psi}(\pi,\omega)$ of $\psi\psi$
states is again flat up to near the upper band edge, where it has a square-root
divergence [Fig.~\ref{fig:fig3}(c)]. Unlike in the case of the $\psi\psi^*$
excitations discussed previously, here the divergence of the density of states
coincides with a zero in the transition rates. In the resulting lineshape, shown
in Fig.~\ref{fig:fig3}(d), the divergence is thus suppressed and converted into
a cusp.

In the limit of zero external magnetic field $(h\to0)$, the $\psi\psi$ continuum
turns into the more familiar two-spinon continuum. The exact two-spinon
lineshape of $S_{- +}(\pi,\omega)$ as previously obtained for $N=\infty$ via algebraic
analysis \cite{KMB+97} does indeed exhibit features very similar to those of
$S_{- +}^{\psi\psi}(\pi,\omega)$ observed here for the first time. In the zero-field
case, however, the power-law singularities are accompanied by logarithmic
corrections.

Our concluding application pertains to the perpendicular spin fluctuations as
described by $S_{+ -}(q,\omega)$, for which we again use the transition rates
(\ref{eq:25}). Whereas the function $S_{- +}(q,\omega)$ dominates the perpendicular
spin fluctuations in weak magnetic fields, it is the function $S_{+ -}(q,\omega)$
that carries most of the spectral weight in strong fields.

A set of dynamically dominant collective excitations is identified which
consists of unbound pairs of $\psi$ and $\psi^*$ quasiparticles as already
encountered in the parallel spin fluctuations.  However, since the perpendicular
and parallel spin fluctuation operators reach these excitations in different
invariant Hilbert subspaces, the spectral boundaries of the $\psi\psi^*$ spectrum
in $S_{+ -}(q,\omega)$ are related to those shown in
Fig.~\ref{fig:fig1}(a) by reflection at the line $q=\pi/2$ \cite{MTBB81,KBM02}.

\begin{figure}[ht]
\centering
\includegraphics[width=6.0cm,angle=-90]{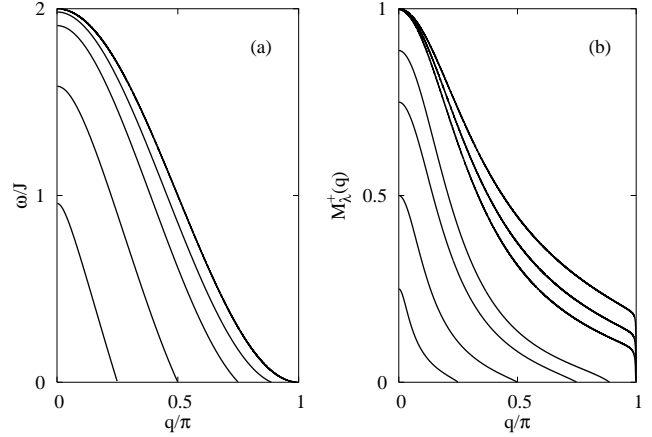}
\caption{(a) Energy-momentum relation at $M_z/N \!=\! 192/1536\!=\!0.125,\; 384/1536\!=\!0.25,\;
  576/1536\!=\!0.375,\; 1364/3072\simeq0.444,\; 9980/20000\!=\! 0.499$ of the $\psi^*$ branch that
  is part of the lower boundary of the $\psi\psi^*$ continuum. (b) Unscaled
  transition rates for the states shown in (a) and those at
  $M_z/N\!=\!9990/20000\!=\!0.4995,\; 9995/20000\!=\!0.49975$.}
\label{fig:fig4}
\end{figure}

One outstanding feature of the spectral-weight distribution $S_{+
  -}^{\psi\psi^*}(q,\omega)$ is the distinct scaling behavior of the transition rates for
the $\psi^*$ branch of the lower boundary \cite{KBM02}. Figure \ref{fig:fig4}(a)
shows the energy-momentum relation of this set of excitations for various values
of $M_z/N$.  This particular branch only exists for $M_z\neq 0$. With $M_z$
increasing from zero, it emerges at $\omega=0, q=0$. The frequency of its
member state at $q=0$ increases proportional to $h$ and the wave number of its
member state at $\omega=0$ increases proportional to $M_z$.  Upon saturation
$(M_z/N\to1/2)$, it turns into the branch of one-magnon states
with dispersion $\omega(q)=J(1+\cos q)$.
In Fig.~\ref{fig:fig4}(b) we plot the unscaled transition rates $M_\lambda^+(q)$ via
(\ref{eq:25}) of these states versus $q$ for the same values of $M_z/N$ as in
panel (a). Also shown are two additional values very close to saturation. The
$N$-dependence at fixed $M_z/N$ of these transition rates is very weak, in
particular at small $q$. We have seen that excitations belonging to a continuum
have transition rates with very different scaling behavior.

The $\psi^*$ transition rate at $q=0$ is exactly known for arbitrary values of
$M_z$: $S_{+ -}(0,\omega) = (2M_z/N)2\pi\delta(\omega-h)$. It represents a resonant mode of
the field-induced magnetization. In the limit $h\to h_S=2J$, the transition rates
become independent of $q$ and carry 100\% of the intensity, $S_{+ -}(q,\omega) =
2\pi\delta(\omega-2J)$, which is well understood in the context of magnon excitations
\cite{MTBB81}. This trend is very slow but clearly visible in the finite-$N$
transition rate data of Fig.~\ref{fig:fig4}(b).

What is perhaps most surprising is that any one of these modes carries a
nonzero fraction of the spectral weight already far below saturation. Hence
they manifest themselves in sharp resonance lines, separate from the adjacent
continuous spectral weight distributions, with intensities that become stronger
as the magnetic field increases.
\begin{acknowledgement}
  Financial support from the DFG Sch\-werpunkt ``Kollektive
    Quantenzust{\"a}nde in elektronischen 1D {\"U}bergangsmetallverbindungen''
  (for M.K.) is gratefully acknowledged.
\end{acknowledgement}

\end{document}